\documentclass[preprint,a4paper]{aastex}
\usepackage{graphics,epsf}
\usepackage{amsmath}                % American Mathematical Society package
\usepackage{amsfonts}               % American Mathematical Society fonts
\usepackage{amssymb}                % American Mathematical Society symbol
\usepackage{epsfig}                 % EPS figures
\usepackage{rotating}

\def \AU{~\rm{AU}}

\def \yr{~\rm{yr}}
\def \Gyr{~\rm{Gyr}}

\begin{document}

\title{THE NUMBER OF PROGENITORS IN THE CORE - DEGENERATE SCENARIO FOR TYPE IA SUPERNOVAE}

\author{Marjan Ilkov\altaffilmark{1} and Noam Soker\altaffilmark{1}}

\altaffiltext{1}{Department of Physics, Technion -- Israel Institute of Technology, Haifa
32000, Israel;  marjan@tx.technion.ac.il; soker@physics.technion.ac.il}

\begin{abstract}
We calculate the expected number of type Ia supernovae (SN Ia) in the core-degenerate (CD) scenario and find it to match observations within the uncertainties
of the code.
In the CD scenario the super-Chandrasekhar mass white dwarf (WD) is formed at the termination
of the common envelope phase from a merger of a WD companion with the hot core of a massive asymptotic giant branch (AGB) star.
We use a simple population synthesis code that avoids the large uncertainties involved in estimating the final orbital separation
of the common envelope evolution. Instead, we assume that systems where the core of the secondary AGB star
is more massive than the WD remnant of the primary star merge at the termination of the common envelope phase.
We also use a simple prescription to count systems that have strong interaction during the AGB phase, but not during the earlier red giant
branch (RGB) phase.
That a very simple population synthesis code that uses the basics of stellar evolution ingredients can match the observed rate
of SN Ia might suggest that the CD-scenario plays a major role in forming SN Ia.
\end{abstract}

% ==========================================================
\section{INTRODUCTION}
\label{sec:intro}
% ==========================================================

There are {{{ {four} }}}  basic theoretical scenarios for the formation of the progenitors of Type Ia supernovae (SNe Ia).
The goal of these scenarios is to form a carbon-oxygen (CO) white dwarf (WD) with a mass close to and above the
Chandrasekhar mass limit $M_{\rm Ch}$.
Such WDs will go through a thermonuclear detonation \citep{HoyleFowler1960} that is observed as an SN Ia.
($i$) In the single degenerate (SD) scenario (e.g., \citealt{Whelan1973}; \citealt{Nomoto1982}; \citealt{Han2004})
a WD grows in mass through accretion from a non-degenerate stellar companion.
However, it seems that the WD-mass increase is very limited, e.g., \cite{Idan2012} for a recent paper and references therein.
{{{ {Ruiter et al. (2011) consider the helium-rich donor scenario (HeRS) to be in a category separate from the canonical SD scenario.
In the HeRS a He-rich star, degenerate or not degenerate, transfers mass to a CO white dwarf which explodes as it
approaches the Chandrasekhar mass (e.g. \citealt{Iben1987}).
Like the SD scenario, the delay time is strongly dependent on evolutionary timescales of the stars.} }}}
($ii$) In the double degenerate (DD) scenario (\citealt{Webbink1984, Iben1984}; see \citealt{vanKerkwijk2010} for a paper on sub-Chandrasekhar mass remnants)
two WDs merge after losing energy and angular momentum through the radiation of gravitational waves \citep{Tutukov1979}.
($iii$) In the core-degenerate (CD) scenario for the formation of SN Ia the Chandrasekhar or super-Chandrasekhar
mass WD is formed at the termination of the common envelope (CE) phase or during the planetary nebula phase,
from a merger of a WD companion with the hot core of a massive
asymptotic giant branch (AGB) star \citep{KashiSoker2011, IlkovSoker2012, Soker2011, Sokeretal2012}.
Observations and theoretical studies cannot teach us yet whether all scenario, or only one or two of these can work
(e.g., \citealt{Livio2001, Maoz2010, Howell2011}).
{{{ { ($iv$) It is likely that a sub-Chandrasekhar mass WD can explode as a SN Ia via the `double-detonation' mechanism
\citep{Woosley1994, Livne1995}.  In this model, a sub-Chandrasekhar mass WD accumulates a layer of helium-rich material
on the surface, which under the right conditions can detonate, leading to a detonation near the center of the CO WD \citep{Fink2010}.
Ruiter et al. (2011) performed a population synthesis study of the double-detonation scenario considering helium-rich donors,
and found a bimodal delay time distribution, the shape depending on whether non-degenerate donors ($<1 \Gyr$)
or degenerate donors ($>1 \Gyr$) dominate. Regarding both the DD and the double-detonation scenarios,
we note that sub-Chandrasekhar SN Ia explosions are supported also by the population of WD binaries
in the solar neighborhood \citep{BadenesMaoz2012}.  } }}}

In this paper we focus on the CD scenario, and to some extend we refer to the DD scenario.
The merger of a WD with the core of an AGB star was studied in the past (\citealt{Sparks1974, Livio2003, Tout2008, Mennekens2010}).
\citet{Livio2003} suggested that the merger of the WD with the AGB core
leads to a SN Ia that occurs at the end of the CE phase or shortly after, and can explain the presence of
hydrogen lines. This idea was further studied recently by \cite{Sokeretal2012}.
In the CD scenario the possibility of a very long time delay (up to $10^{10}$~yr) is considered as well
\citep{IlkovSoker2012}.
{{{ {However, if that delay mechanism does not work, then the CD scenario can explain SN Ia only in star forming galaxies.} }}}
\cite{Mennekens2010} {{{ { and \cite{Ruiter2008} } }}}  find in their population synthesis calculations that many
systems end in a core-WD merger process, but they did not consider these to be SN Ia.
Because of its rapid rotation  (e.g., \citealt{Anand1965}; \citealt{Ostriker1968}; \citealt{Uenishi2003}; \citealt{Yoon2005}),
and possibly very strong central magnetic fields (e.g., \citealt{Kundu2012, Garcia-Berro2012}), the super-Chandrasekhar WD does not explode.

There are two key ingredients in the CD scenario, in addition to the common condition that the remnant mass be
$\ga 1.4 M_\odot$.
(1) The merger should occur while the core is still large, hence hot. This limits the merger to occur within $\sim 10^5 \yr$
after the common envelope phase. \citet{KashiSoker2011} showed that this condition can be met when the AGB star is massive,
and some of the ejected CE gas might fall back (see also \citealt{Soker2012}).
(2) The delay between merger and explosion should be up to $\sim 10^{10} \yr$ if this scenario is to account
for SNe Ia in old-stellar populations, in addition to SNe Ia in young stellar populations.
WDs with delay times have masses in the range $\sim 1.4-1.5 M_\odot$, and indeed can have a long delay
{{{ { (Jorge Rueda, 2012, private communication).} }}}
In previous papers another ingredient was used. It asserted that the hot core is more massive than the companion cold WD.
This is true in most cases, but not in all. In some rare cases this not need be the case \citep{Sokeretal2012}, and there is a
short delay to explosion.

These ingredients involve physical processes much different from those in the DD scenario, and hence
make the CD a distinguished scenario, rather than a branch of the DD scenario.
The merger of the core while it is still hot might prevent an early ignition of carbon \citep{Yoon2007}, which
is one of the theoretical problems of the DD scenario (e.g., \citealt{SaioNomoto2004}).
As for the delay time, in the CD scenario it is due to the spinning-down time of the WD set by magneto-dipole radiation
\citep{IlkovSoker2012}, while in the DD scenario the delay is due to the spiraling-in time due to gravitational radiation.

Stabilizing rapidly rotating super-Chandrasekhar WDs is a delicate matter (e.g., \citealt{Boshkayev2012}, \citealt{Yoon2004},
\citealt{Chen2009}, and \citealt{Hachisuetal2012a, Hachisuetal2012b} in the SD degenerate scenario).
The strong magnetic fields required in the present model most likely will enforce a rigid rotation within a short
time scale due the WD being a perfect conductor. The critical mass of rigidly rotating WDs is $1.48 M_\odot$
(\citealt{Yoon2004} and references therein).
This implies that WDs more massive than $1.48 M_\odot$ will explode in a relatively short time.
The similarity of most SN Ia suggests that their progenitors indeed come from a narrow mass range.
This is $\sim 1.4-1.48 M_\odot$ in the CD scenario.
This property of the magneto-dipole radiation torque spinning-down mechanism,
{{{ {that only WDs with $M_{\rm WD} \la 1.48 M_\odot$ can
slow down on a very long time scale (e.g., Jorge Rueda, 2012, private communication), }  }}}
explains the finding that SNe Ia in older populations are less luminous (e.g., \citealt{Howell2001, Smith2011}),
and that very massive SN Ia progenitors occur in galaxies where star formation is expected, i.e., spirals and irregulars \citep{Scalzoetal2012}.

In the present paper we estimate the number of SNe Ia from the CD scenario.
Because of several large uncertainties, we perform a crude estimate of the SN Ia progenitor formation rate.
In section \ref{sec:coincident} we point to an interesting coincident of the Chandrasekhar mass with
another mass in the CD and DD scenarios.
The assumptions and set up of the population synthesis are given in section \ref{sec:assumptions}.
The results on the formation rate are given section \ref{sec:results},
together with a comparison with observations, including with core-collapse supernovae (CCSNe) rate.
Our summary is in section \ref{sec:summary}.

% ==========================================================
\section{THE $\sim 1.35 M_\odot$ COINCIDENCE }
\label{sec:coincident}
% ==========================================================

The electron degeneracy pressure operating in two different stellar phases leads to an interesting coincidence
in the frame of the scenario discussed here.
Relativistic electron degeneracy pressure sets the Chandrasekhar mass limit of $M_{\rm Ch} = 1.4 M_\odot$.
Non-relativistic electron degeneracy pressure operates in the core of low mass red giant branch (RGB) stars.
These stars reach a relatively large radius on the RGB.
The combined mass of the two WD remnants of a binary system where both stars develop a degenerate core on the RGB is limited by
$\sim 1.3 M_\odot$. When evolutionary considerations are added (see below) this mass is very close to $M_{\rm Ch}$.

\begin{figure}
\includegraphics[scale=0.5]{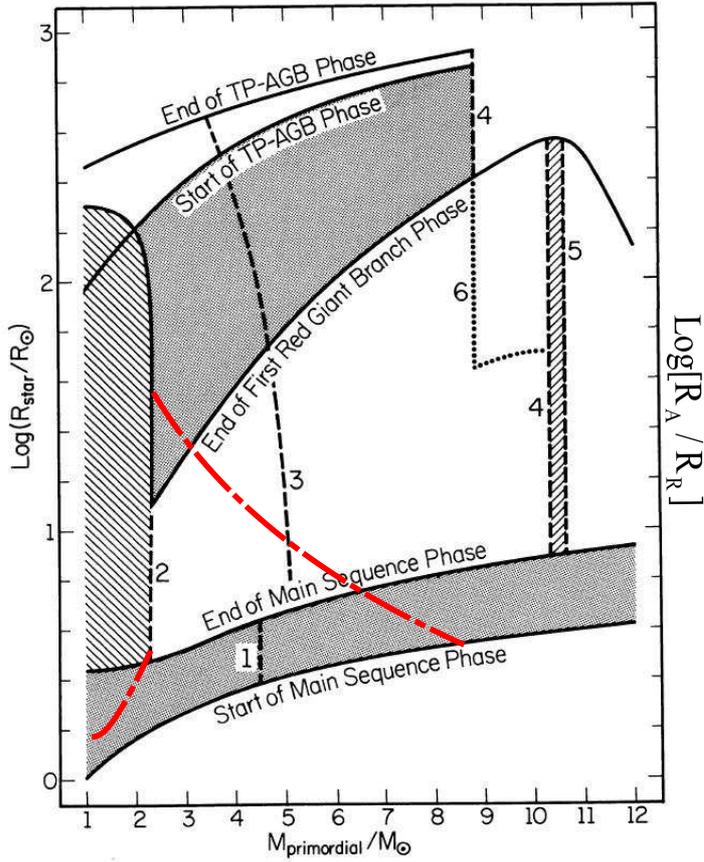}
        \caption{ The maximum radius $R_R$ on the red giant branch (RGB) and the one $R_A$ on asymptotic giant branch (AGB)
         attained by starts as function of their main sequence (MS) mass, as calculated by \cite{Iben1985}.
         We added two thick dashed-dotted lines that are the quantity $\log (R_A/R_R)$ as given by our approximate equations (\ref{eq:RARR1}) and (\ref{eq:RARR2}) for the
         two mass ranges, respectively.
            }
   \label{fig:fig1}
     \end{figure}

Stars that have a main sequence (MS) mass of $M_{\rm MS} \la 2.3 M_\odot$ develop a degenerate core and reach a
relatively large radius on the RGB, not much smaller than the maximum radius they achieve on the asymptotic giant branch (AGB).
Stars with $M_{MS} \ga 2.3 M_\odot$, on the other hand, reach a much larger radius on the AGB relative to the radius
they achieve on the RGB.
The following approximation (up to $20\%$) can be fitted to the results of \citet{Iben1985}(their fig. 31) for the
ratio of the maximum radius on the AGB to that on the RGB as function of the MS mass
\begin{equation}
\log (R_A/R_R) = 3.7 \log^2 (M/M_\odot) - 0.37 \log (M/M_\odot) + 0.16, \qquad M \la 2.25 M_\odot,
\label{eq:RARR1}
\end{equation}
and
\begin{equation}
\log (R_A/R_R) = 2.2  - 1.8 \log (M/M_\odot), \qquad M \ga 2.35 M_\odot.
\label{eq:RARR2}
\end{equation}
The results of \cite{Iben1985} together with the fitting formula are plotted in Figure \ref{fig:fig1}.
This implies that for stars with $M_{\rm MS} \ga 2.3 M_\odot$ in binary systems there is a large range of orbital separations
where strong binary interaction occurs during the AGB phase but not during the RGB phase of the star.
This condition is required to form a CO WD merger remnant.
% FFFFFFFFFFFFFFFFFFFFFFFFFFFFFFFFFFFFFFFFFFFFFFFFFFF
% FFFFFFFFFFFFFFFFFFFFFFFFFFFFFFFFFFFFFFFFFFFFFFFFFFF

It so happens that a star of MS mass of $M_{\rm MS} \simeq 2.3$ that evolves without any disturbances forms a WD of mass
$M_{\rm WD} (2.3) \simeq 0.65 M_\odot$ (e.g., \citealt{Catalanetal2008, Catalanetal2009}).
This leads to the following argument.
Consider binary systems where for both stars there is a large range of the binary orbital separation
for a strong binary interaction to occur on the AGB, but not on the RGB.
Namely, both stars have $M_{MS} > 2.3 M_\odot$ when they leave the MS.
The secondary can start the MS with $M_{MS0} < 2.3 M_\odot$, but then accretes mass from the primary
star to the point where its mass is $M_{\rm 2new}>2.3 M_\odot$. We use the condition that after mass transfer the secondary mass is  $M_{\rm 2new} > 2.3 M_\odot$.
We also use the condition $M_2 > 1 M_\odot$ on the initial mass of the secondary star.
This ensures that in most cases the binary system does not reach the Darwin instability during the AGB phase of the primary star,
and so there is no CE phase at this stage.
Namely, we consider systems where the secondary (initially less massive star) is massive enough to bring the primary envelope
to synchronization with the orbital motion, and a CE phase before the secondary has developed a CO core is prevented .
The primary forms a WD of mass $M_{WD1} \ga 0.65 M_\odot$.

Later in the evolution the secondary evolves and interaction takes place only at late AGB phases.
The WD remnant of the primary star cannot bring the secondary envelope to synchronization, and the binary system
enters a CE phase.
However, as this occurs late on the AGB when the core is a CO core and is massive enough, $M_{\rm core} \ga 0.6 M_\odot$.
Despite the large uncertainties in the binary interaction, systems where the combined remnants mass is
$M_{\rm WD1} + M_{\rm core} \ga 1.4 M_\odot$ are very likely to
be descendant of two stars that each left the main sequence with a mass of $M_{\rm MS} > 2.3 M_\odot$.

On the other hand, WD binary systems with $M_{\rm WD1} + M_{\rm core} \la 1.3 M_\odot$ are likely to be
descendant of systems where at least one of the two stars had a mass of $M_{MS} \la 2.3 M_\odot$
and a strong tidal interaction took place on the RGB. This would further prevent the WD from achieving a high mass.

The WDs remnants, or a WD and the core of the secondary, of two stars in a binary system that
are likely to have strong interaction on the AGB but not on the RGB, is about equal to the Chandrasekhar mass limit.
The condition is that both stars have $M_{MS} > 2.3 M_\odot$ when they leave the MS, and the orbital separation be in the
right range. The orbital separation range for forming a SN Ia progenitor is not simple to estimate, as it requires knowledge of the tidal interaction.
In any case, in the next section we will use the logarithmic range $\Delta \equiv \log (R_A) - \log (R_R)$
in estimating the number of systems that potentially can lead to SNe Ia.

When a low mass secondary star, i.e., $M_{MS} < 2.3 M_\odot$, evolves, it has a lower envelope mass.
A WD companion (the remnant of the primary) is more likely to survive the CE phase.
The remnant is a close WD binary system with a total mass of $M_{t} \la 1.3 M_\odot$.
Those WD binaries with $M_{t} > 1.4 M_\odot$ are likely to come from systems where the secondary had $M_{MS} > 2.3 M_\odot$,
and the CE is more likely to end with a merger \citep{Soker2012}.
This is because the outcome of a CE phase is sensitive to the envelope mass. A small increase in the envelope
mass above the combined masses of the core and WD companion can lead to a merger event at the termination of the CE phase \citep{Soker2012}.
To have a core mass of $M_{\rm core} > 0.7 M_\odot$ the secondary MS mass should be $M_{\rm MS} \ga 3 M_\odot$.
The envelope mass is then $\sim 2.3 M_\odot$ which is 3 times and more the companion WD mass.

This general discussion might explain the finding of \citet{BadenesMaoz2012} that the super-Chandrasekhar merger rate of
WD binary systems in the solar neighborhood, $\sim 10^{-14} \yr^{-1} M_\odot^{-1}$, is only $\sim 10\%$ of the SN Ia rate found by \citet{Li2011}.
\citet{BadenesMaoz2012} suggest that the explanation is that sub-Chandrasekhar mergers lead to SNe Ia.
To account for the observed rate they require SNe Ia to occur in merger of WD with a total mass of as low as $\sim 1.1 M_\odot$.
{ {{{ \cite{Ruiter2011} find in their population synthesis that theoretically there are potentially enough sub-Chandrasekhar progenitors to account
for the rates of SNe Ia. } }}}
Instead, we suggest that many single compact-magnetic super-Chandrasekhar WDs explode as SNe Ia. They are the remnants of
core-WD merger at the termination of the CE phase.
The recent population synthesis of \cite{Garcia-Berro2012} show that many more mergers occur during the AGB phase than of two WDs (about a factor of 8).

% ==========================================================
\section{ASSUMPTION AND SET UP}
\label{sec:assumptions}
% ==========================================================
In this section we list the assumptions and set up of the population synthesis calculations.
Because of the large uncertainties in the CE evolution and the controversial $\alpha$-CE prescription \citep{Soker2012},
we limit our goal to show that the number of SN Ia can be explained with the CD and DD scenario.
Previous population synthesis of the the DD scenario (e.g., \citealt{ Yungelson2005, Ruiter2009, Mennekens2010, WangHan2012})
are summarized by \cite{Nelemansetal2012} and \cite{WangHan2012}.
{ {{{ The more recent population synthesis study conducted by \cite{Toonen2012} shows that the expected total number SN Ia progenitors in the DD scenario
is a factor of $\sim 7-12$ times lower than observed. Even if the fraction of binary is taken to be $70\%$ instead of $50 \%$, the contribution of
the DD scenario is $\la 20\%$. This is compatible with the observational finding of \cite{BadenesMaoz2012}. }}} }
With the CE prescription used in the context of the DD scenario, and for $50 \%$ of systems in binaries,
the results span the range of $0.2-0.6$~SN Ia per $1000 M_\odot$ star formation,
which is below half the observed value.
As we will show below, the CD scenario does much better and can account for the observed value.

There are three basic differences between our calculations and those summarized  by \cite{Nelemansetal2012} and \cite{WangHan2012}.
($i$) We consider the merger of the WD companion with the core, in particular in cases where the
WD mass is less than the core mass, to be of very high probability \citep{Soker2012}.
($ii$) We use a simple prescription for the relevant binary orbital separation that might allow stronger interaction on the AGB (see more in section \ref{sec:summary}).
($iii$) Our usage of a high efficiency mass transfer ($\eta \ga 0.5$ below) is more than what other population synthesis codes
use/find.
This effect might be the largest of the three effects listed here.
In that regards we note that our results disagree with those of \cite{MengYang2012} regarding the CD scenario, both quantitatively and
qualitatively. It is possible that most of the differences result from our usage of this efficient mass transfer process.

% ==========================================================
\subsection{Binary evolution}
\label{subsec:evolution}
% ==========================================================
The ingredients of the population synthesis code are as follows.
\newline
(1) {\it Initial mass functio (IMF).} We use the IMF of \cite{Kroupa1993} where
 \begin{equation}
   f(m) = \left\{
     \begin{array}{lr}
       0.035 m^{-1.3}, \quad {\rm for} \quad   0.08 < m < 0.5M_\odot \\
       0.019 m^{-2.2}, \quad {\rm for} \quad  0.5 < m < 1.0M_\odot \\
       0.019 m^{-2.7}, \quad {\rm for} \quad 1.0 < m < \infty ,  \\
     \end{array}
   \right.
      \label{eq:IMF1}
\end{equation}
where $m \equiv M/M_\odot$.
In our numerical monte-carlo code we take the formula as suggested by \cite{Kroupa1993}
 \begin{equation}
M =  0.08 + \frac{(0.19X^{1.55} + 0.050X^{0.6})}{(1 - X)^{0.58}},
      \label{eq:IMF2}
\end{equation}
where $X$ is a uniformly distributed random number in the interval [0, 1].

(2) {\it Upper mass limit of $M<7 M_\odot$.} From these stars we choose the ones in the mass range $2.3  M_\odot < M_1 <  7M_\odot$.
These are the primary stars in what follows.
The secondary in each binary system is taken to have an equal probability to be in the mass range  $0.08 M_\odot < M_2 < M_1$.
These masses take into consideration that primaries with a mass
of $M_1>7 M_\odot$ either explode as a core collapse supernova (CCSN) or form a NeMgO WD.

(3) {\it Lower mass limits.} Once $M_1$ evolves it might interact with the companion.
To form a CO WD we require that a strong tidal interaction takes
place on the AGB phase of the both stars, but not during the RGB phase.
For the statistical purposes of this study we are not interested in the value of the orbital
separation where tidal interaction takes place, but only in the ratio of that orbital interaction on the AGB to that on the RGB
(see section \ref{sec:coincident}).
Crudely, this ratio is the ratio of the maximum radius of the star on the AGB, $R_A$, to that on the RGB, $R_R$, as given by equation
(\ref{eq:RARR2}) taken from \cite{Iben1985}. For $M<2.35 M_\odot$ this ratio is very close to 1, and we ignore systems with $M_1<2.3 M_\odot$.
After mass transfer takes place we make sure that $M_{\rm 2new}>2.3 M_\odot$.
We ignore systems with low initial secondary mass of $M_2<1 M_\odot$.
Low mass secondary stars that are potentially in the relevant orbital separation will not be able to bring the primary envelope
to synchronization with the orbital motion, and will enter a CE phase too early.

(4) {\it Mass transfer.} The strong interaction during the AGB will lead to mass transfer from the primary to the secondary.
We consider here a channel where the mass transfer from the primary to the secondary avoids a CE phase.
A CE phase occurs only when the secondary becomes a giant.
\cite{Mennekens2010} find this channel to be more common than the two-CE phases channel in their
population synthesis study of the DD scenario (some details and examples of the evolutionary channels can be found there).
We parameterize the poorly known mass transfer process as follows.
We take the new mass of the secondary after the primary has gone through the AGB phase and turned into a WD
of mass $M_{\rm WD1}$ as
\begin{equation}
M_{\rm 2new} = M_2 + \eta (M_1 - M_{\rm WD1}),
\label{eq:eta1}
\end{equation}
where $0 \le \eta \le 1$. From a more accurate population synthesis, using the code described by \cite{Garcia-Berro2012},
of the mass transfer process from the evolved primary to the secondary,
the average value of $\eta$ is found to be $\bar \eta \simeq  0.9$ (Camacho, J. 2012, private communication).
We will take $\eta=0.8$ as our standard case.

(5) {\it WD Masses.} The mass of the WD that emerges from the evolution of each star is taken according to
\cite{Catalanetal2008} as
 \begin{displaymath}
   M_{\rm WD1} = \left\{
     \begin{array}{lr}
       (0.096 \pm 0.005) M_1 + 0.429 \pm 0.015, & M_1<2.7M_\odot \\
       (0.137 \pm 0.007) M_1 + 0.318 \pm 0.018, & M_1>2.7M_\odot ,
     \end{array}
   \right.
\end{displaymath}
where for the WD remnant of the secondary $M_1$ is replaced by $M_{\rm 2 new}$.

(6) {\it Binary fraction.} Our standard fraction of binaries for these massive stars is taken to be in the range $f_b=0.5-0.65$, where this
fraction includes secondaries in the mass range $0.08 M_\odot-M_1$.
\cite{Nelemansetal2012} use $f_B=0.5$, but our condition of $M_1 > 2.3 M_\odot$ and $M_2 > 1 M_\odot$ takes more massive stars, where the binary fraction can be somewhat
larger that $f_b=0.5$, {{{{ up to $\sim 70 \%$ (see summary of observations in fig. 12 of \cite{Raghavanetal2010}. }}}}

% ==========================================================
\subsection{Counting SNIa Progenitors}
\label{subsec:counting}
% ==========================================================

The criteria to count a system as a potential SN Ia progenitor are as follows.
\newline
(1) {\it Mass considerations.} For the SN Ia progenitors we count only systems with
\begin{equation}
M_{\rm WD1} + M_{\rm core} > 1.4 M_\odot.
\label{eq:masst}
\end{equation}
In most works of this type the final orbital separation is calculated using the controversial CE $\alpha$-prescription.
We here follow \cite{Soker2012} and assume that most cases where the AGB envelope is massive enough will end with merger.
This is particularly the case when the WD remnant of the primary star has a lower mass than the core of the secondary,
namely, $M_{\rm WD1} < M_{\rm core}$.
These systems will be counted separately as the likely progenitors of SN Ia in the CD scenario.

(2) {\it Orbital separation considerations.}
We consider potential progenitors of WD-core or WD-WD merger to be systems (namely, systems where the WD enters the envelope of the secondary
during the AGB phase of the secondary star) that have strong tidal interaction on the AGB but not on the RGB.
The relevant initial orbital separation will be in the range $C_o R_R <a < C_o R_A$, where $R_R$ and $R_A$ are the maximum
radius of the star on the RGB and AGB, respectively. The factor $C_o \sim 2-5$ is determined by the tidal forces.
The relevant range for our potential progenitors in the logarithm of the orbital separation scale is therefore
$\log (C_o R_A)- \log( C_o R_R)= \log (R_A/R_R)$.
We take the initial orbital separation of the binary population of massive stars created by the numerical code to span a range of 4.5 orders of magnitude
(from $a_{\rm min} \sim 10 R_\odot$ to $a_{\rm max} \sim 1500 \AU$) with an equal probability in the logarithmic of the orbital separation.
Accordingly, the probability of a binary system to have the initial relevant orbital separation is
\begin{equation}
x_i = \frac{\log (C_o R_A)-\log (C_o R_R)} {\log a_{\rm max}- \log a_{\rm min}}  \simeq
\frac{1}{4.5} \log \frac {R_A}{R_R} =\frac{1}{4.5} \left( 2.2 - 1.8 \log M_{\rm star} \right),
\label{eq:xi1}
\end{equation}
where equation \ref{eq:RARR2} based on \cite{Iben1985} have been used in the last equality.
Instead of calculating for each of the two stars, in the present study it is adequate to take the average mass for this criterion
$M_{\rm star} =0.5(M_1+M_2)$.

(3) {\it Progenitors of CCSNe.}
Systems that have $M_1 > 8.5 M_\odot$ are counted as having one CCSN, while if $M_2> 8.5 M_\odot$, the system produces two CCSNe
\citep{Gilmore2004}. This number will be used to find the ratio between the number of SN Ia and that of CCSN,
and compare with observations, e.g., as given by \cite{Li2011}.

(4) {\it Avoiding NeMgO WDs.} Merger remnants that have a large fraction of neon and magnesium cannot explode. We therefore avoid NeMgO WDs by
not counting systems where either $M_{\rm WD1} > 1.1 M_\odot$ or $M_{\rm core} > 1.1 M_\odot$.

% ==========================================================
\section{RESULTS}
\label{sec:results}
% ==========================================================
% ==========================================================
\subsection{The number of potential SN Ia progenitors}
\label{subsec:number}
% ==========================================================
We consider the parameters range of the model to be $f_b \simeq 0.5-0.65$ and $\eta \simeq 0.5-0.9$,
for the fraction of binary systems and mass transfer parameter, respectively.
The number of systems obtained under the counting conditions listed in section \ref{subsec:counting} are presented in Table \ref{tab:Table1}.
The first panel gives the number of potential progenitors of SN Ia per $1000 M_\odot$ of stars formed, while the second panel
gives the number of potential progenitors relative to the number of CCSNe.
We also give the numbers only for systems that in addition to the criteria given in section \ref{sec:assumptions}, also obey the
condition that the WD mass is lower than the core mass, $M_{\rm WD1} < M_{\rm core}$.
Such systems are more likely to merge at the termination of the CE phase.
% TTTTTTTTTTTTTTTTTTTTTTTTTTTTTTTTTTTTTTTTTTTTTTTTTTTTTTTTTTTTTTTTTTTTTTTT
\begin{table}[]
  \centering
  \caption{The calculated number of systems entering the CE phase according to the counting criteria listed
  in section \ref{subsec:counting}.   Upper panel: number of systems per $1000 M_\odot$ of star formation.
  The observed value is from \cite{Maoz2012}.
  Second panel: The calculated ratio of the number of potential SN Ia progenitors to core collapse SNe (CCSN).
  The observed range is from \cite{Maoz2011}, \cite{Li2011} and \cite{Melinder2012}.
  The two rows labelled with superscript $\ast$ in both panels are for systems with $M_{\rm WD1} < M_{\rm core}$, that are more
  likely to suffer core-WD merger and have long delay to explosion.  }
    \begin{tabular}{rrr}
          &              &  \\
    \hline
\multicolumn{3}{c}{SNIa per $1000 M_\odot$} (Observed value $1-2$)   \\
        \hline   & \multicolumn{1}{c}{$\eta =0.5$} & \multicolumn{1}{c}{$\eta =0.8$} \\
 \hline
 \multicolumn{1}{c}{$fb=0.5$}         & \multicolumn{1}{c}{2.0}  & \multicolumn{1}{c}{2.3} \\
 \multicolumn{1}{c}{$fb=0.5^\ast$}    & \multicolumn{1}{c}{1.2}  & \multicolumn{1}{c}{2.0} \\
 \hline
 \multicolumn{1}{c}{$fb=0.65$}        & \multicolumn{1}{c}{2.4}  & \multicolumn{1}{c}{2.8} \\
 \multicolumn{1}{c}{$fb=0.65^{\ast}$} & \multicolumn{1}{c}{1.4}  & \multicolumn{1}{c}{2.4} \\
    \hline
          &              &  \\
          \\
     \\
   \hline    \multicolumn{3}{c}{SNIa/CCSN}  (Observed value $0.2-0.4$)        \\
    \hline     & \multicolumn{1}{c}{$\eta =0.5$} & \multicolumn{1}{c}{$\eta =0.8$} \\
    \hline
\multicolumn{1}{c}{$fb=0.5$}          & \multicolumn{1}{c}{0.44}  & \multicolumn{1}{c}{0.51} \\
\multicolumn{1}{c}{$fb=0.5^\ast$}     & \multicolumn{1}{c}{0.26}  & \multicolumn{1}{c}{0.44} \\
   \hline
\multicolumn{1}{c}{$fb=0.65$}         & \multicolumn{1}{c}{0.55} & \multicolumn{1}{c}{0.65} \\
\multicolumn{1}{c}{$fb=0.65^{\ast}$}  & \multicolumn{1}{c}{0.32} & \multicolumn{1}{c}{0.55} \\
   \hline
          &       &  \\
     \end{tabular}
  \label{tab:Table1}
\end{table}
% TTTTTTTTTTTTTTTTTTTTTTTTTTTTTTTTTTTTTTTTTTTTTTTTTTTTTTTTTTTTTTTTTTTTTTTT

The number of potential progenitors for the standard parameters and with the condition of
$M_{\rm WD1} < M_{\rm core}$, is $\sim 1-2$ times the observed value.
There are several processes that can change the number of potential SN Ia progenitors in our modelling, but not by much.
($i$) Some cases with $M_{\rm WD1} > M_{\rm core}$ can also lead to SN Ia, although
not many such progenitors are expected \citep{Sokeretal2012}.
On the other hand, we might overestimate the number of progenitors as during the merger process some
mass might be lost, hence some systems will become sub-critical in their final merger product mass.
As well, in some cases the primary WD enters the envelope before the core reaches its maximum mass on the AGB, $M_{\rm core}$.
Hence, some of the systems counted here as having a total mass above the Chandrasekhar mass, might actually be below the critical mass.
($ii$) The binary fraction can be larger than $f_b=0.5$, and $f_B \simeq 0.65$ might be more appropriate
{ {{{ for the B-type stars considered here (e.g., see summary by \cite{Raghavanetal2010}). }}} }
On the other hand, in some systems a tertiary star might disturb the evolution and prevents a SN Ia from occurring.
($iii$) A value of $\eta=0.9$ is more appropriate. This will increase the number of progenitors.
On the other hand, some unstable phases in binary evolution might lead to extra mass loss
that is not considered in population synthesis calculations. As well, some systems might enter a CE evolution during the
AGB phase of the primary. This might prevents the secondary star from developing a CO core later on.

Overall, we consider it very satisfactory to the CD scenario that with a very simple modelling we find the number of SN Ia
to be slightly above the observed value.
This margin comes for the over-counting mentioned above.
More sophisticated population synthesis calculations that take mergers into account should
be performed to derive more accurate values for the expected number of SN Ia in the CD scenario.
{{{ {As well, the time delay should be calculated. Here we only compare the total number (time integrated) of SN Ia,
and not the evolution of rate with time. The model should eventually reproduce the delay time distribution of SN Ia if it to account for most SN Ia.} }}}
If the merging core-WD events found by
\cite{Mennekens2010} did not consider the many core-WD merger events they found to be SN Ia.
If these events are considered to be SN Ia, then their calculations yield a similar rate to the one
we obtain here under the same conditions (N. Mennekens 2012, private communication).

% ==========================================================
\subsection{Properties of the merging core-WD systems}
\label{subsec:remnant}
% ==========================================================
 We present the distribution in some properties of the WD-core remnants.
All graphs in this subsection are for systems that obey the conditions of section \ref{subsec:counting} and for the mass
transfer parameter of $\eta=0.8$.
{{{ {In these graphs 500 systems correspond to $0.073$ systems per $1000 M_\odot$ of star formation for $f_B=1$ (all stars are binary).
For other values of $f_B$ the number should be scaled accordingly.} }}}

\begin{figure}[b]
% \vspace*{-2.0 cm}
\begin{center}
 \includegraphics[scale=0.7]{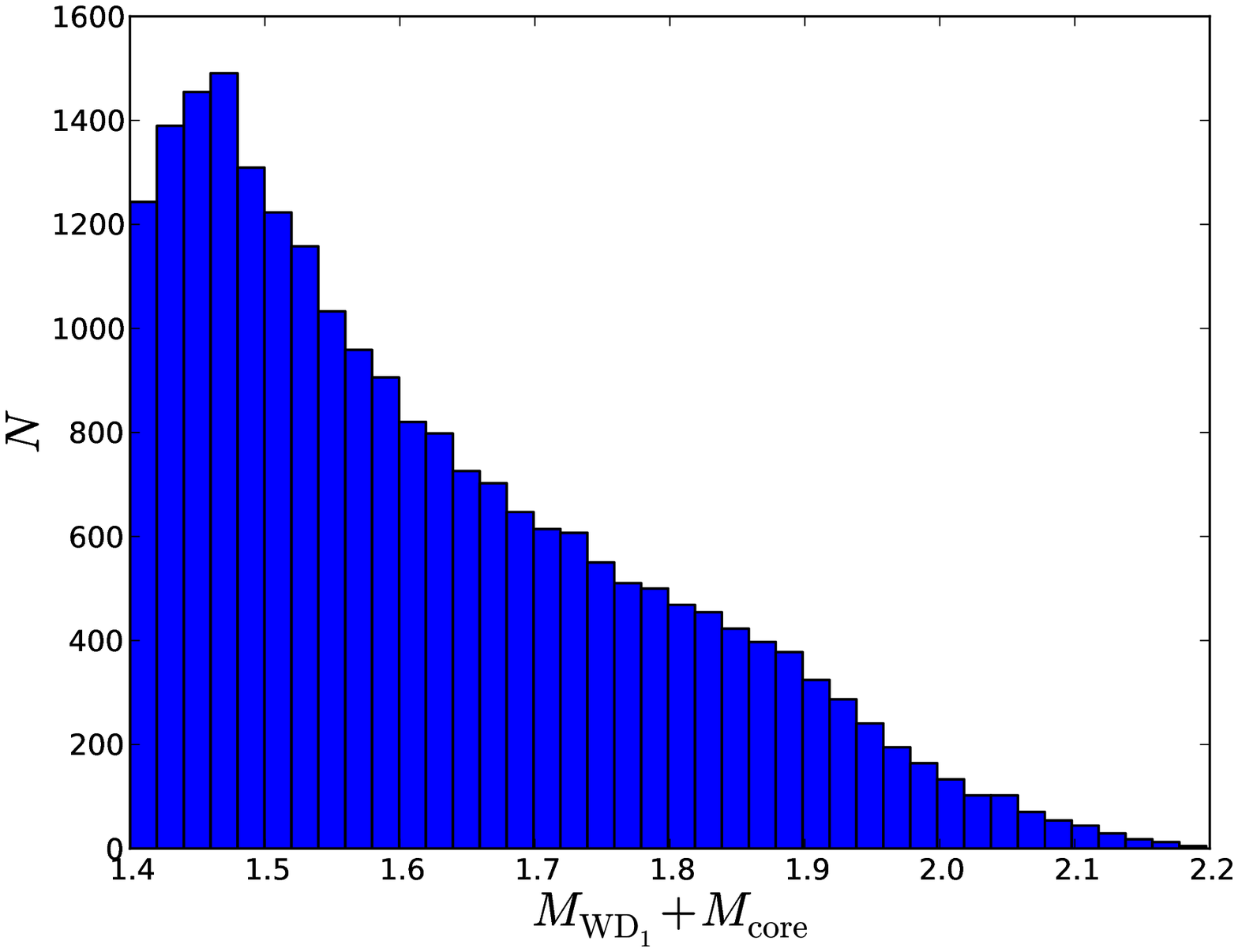}
% \vspace*{-1.0 cm}
\caption{Histogram of the combined masses of the two remnants $M_t=M_{\rm WD1} + M_{\rm core}$.
Only systems with $M_{\rm WD1} < M_{\rm core}$ are included in this figure.
On average, these are more likely to merge at the end of the CE phase.  }
           \label{fig:fightmass}
\end{center}
\end{figure}

\begin{figure}[b]
% \vspace*{-2.0 cm}
\begin{center}
 \includegraphics[scale=0.5]{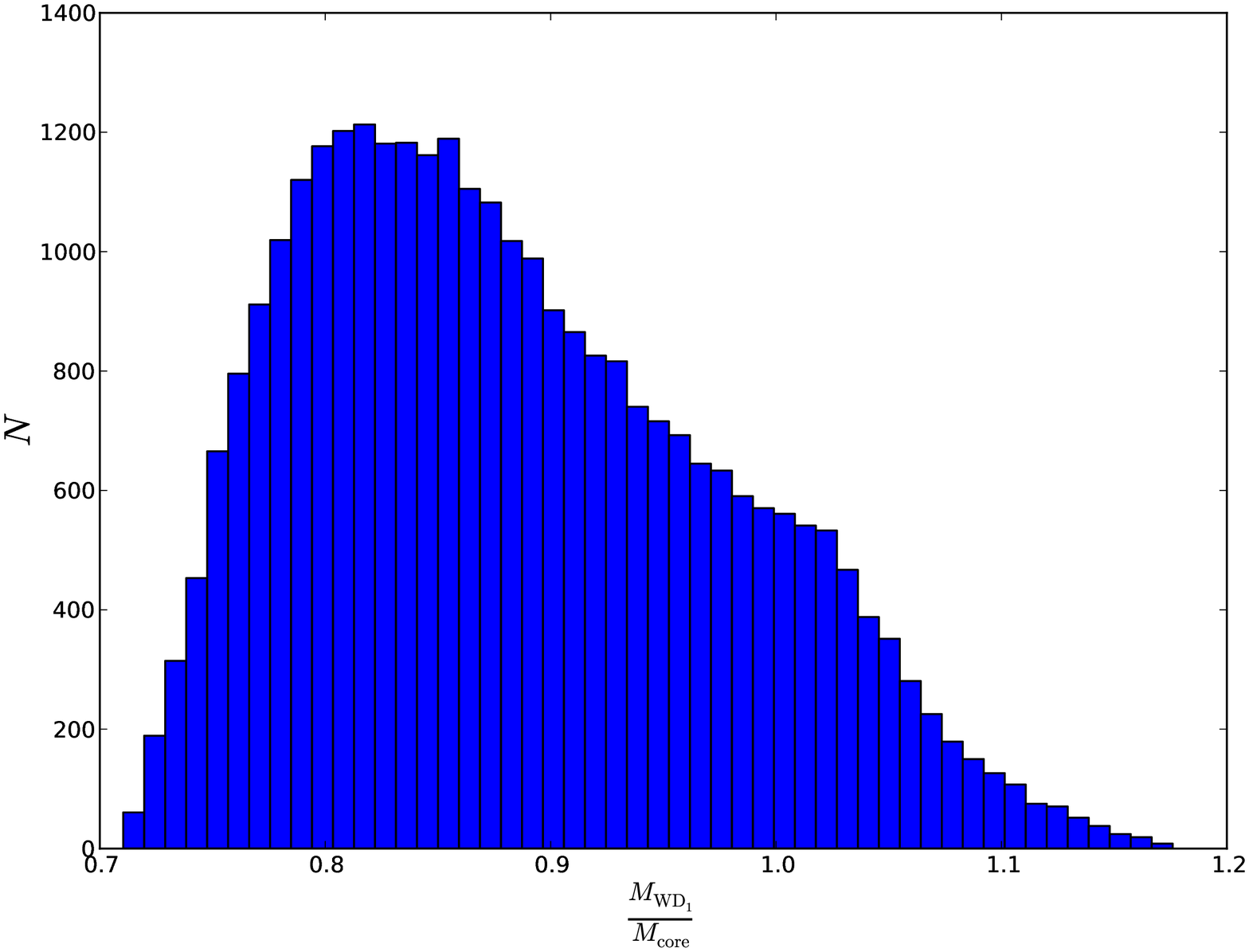}
% \vspace*{-1.0 cm}
        \caption{Histogram of the WD to core mass ratio, $M_{\rm WD1}/M_{\rm core}$ for the systems
        studied in this section.
 Note that a large number of systems have the AGB core mass larger than the mass of the WD during the final CE phase. }
           \label{fig:figratio12}
\end{center}
\end{figure}

In Figure \ref{fig:fightmass} we present the mass histogram of the total mass $M_t=M_{\rm WD1} + M_{\rm core}$ for systems with
$M_{\rm WD1} < M_{\rm core}$. In most of these systems the cold primary WD is expected to be destructed on the core. A long time delay
to explosion is likely.
We emphasize that the merger process might eject some mass (e.g., \citealt{Han1999, ChenX2012}), and the final merger product mass might be smaller.
Due to mass loss during the merger process, we expect that more systems will end in the mass range $1.4-1.5 M_\odot$,
where they expect to have a very long time delay to explosion
{{{ { (up to $\sim 10^{10} \yr$), } }}}
long after the star formation episode ceased.
Systems with mass much above the Chandrasekhar mass limit, but where the WD is destructed on the core,
will have shorter time delay, but longer that the time the ejected envelope disperses in the ISM, $\sim 10^5 \yr$,
and are expected to explode in star-forming regions.
Systems where $M_{\rm WD1} \ga M_{\rm core}$ are less likely to merge. If they do merge, then the core might be destructed on the cooler WD,
and explosion might occur within tens of years, leading to the presence of hydrogen lines \citep{Sokeretal2012}.
The histogram of the mass ratio for any remnants mass ratio is given in Figure \ref {fig:figratio12}.
% FFFFFFFFFFFFFFFFFFFFFFFFFFFFFFFFFFFFFFFFFFFFFFFFFFFFFF
% FFFFFFFFFFFFFFFFFFFFFFFFFFFFFFFFFFFFFFFFFFFFFFFFFFFFFF
% FFFFFFFFFFFFFFFFFFFFFFFFFFFFFFFFFFFFFFFFFFFFFFFFFFFFFF
% FFFFFFFFFFFFFFFFFFFFFFFFFFFFFFFFFFFFFFFFFFFFFFFFFFFFFF

\begin{figure}[b]
% \vspace*{-2.0 cm}
\begin{center}
\includegraphics[scale=0.6]{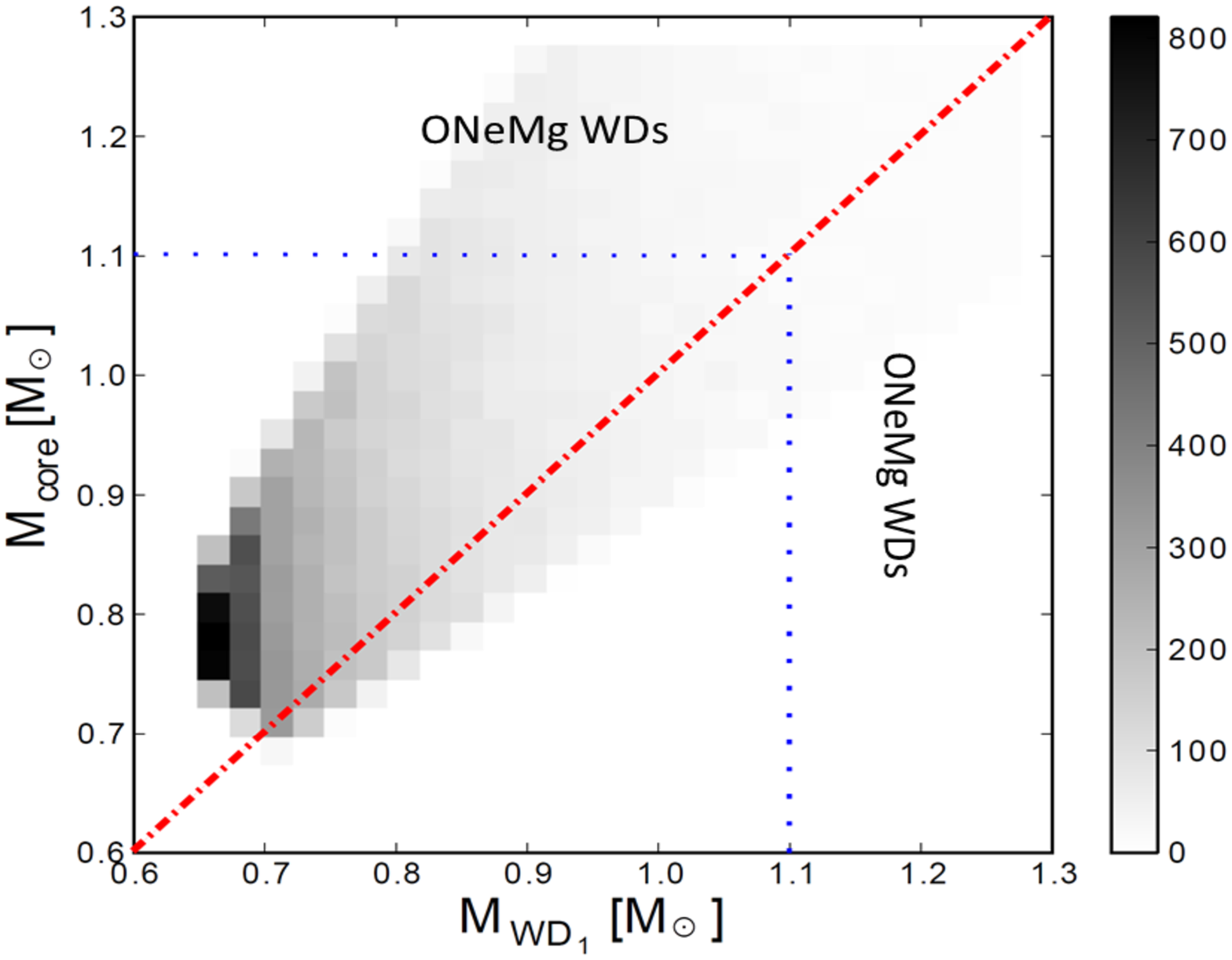}
% \vspace*{-1.0 cm}
        \caption{Density of number of systems in the $M_{\rm WD1}-M_{\rm core}$ plane.
        The scale is the number of systems in each square (of $0.025\times 0.025 M_\odot^2$).
        Only in this figure we have systems with $M_{\rm WD1} > 1.1 M_\odot$ and/or $M_{\rm core} > 1.1 M_\odot$ included.
        This is for comparison purposes only,
as systems where either $M_{\rm WD1} > 1.1 M_\odot$ or $M_{\rm core} > 1.1 M_\odot$ are not counted as SN Ia progenitors.}
           \label{fig:figden12}
\end{center}
\end{figure}

\begin{figure}[b]
% \vspace*{-2.0 cm}
\begin{center}
\includegraphics[scale=0.6]{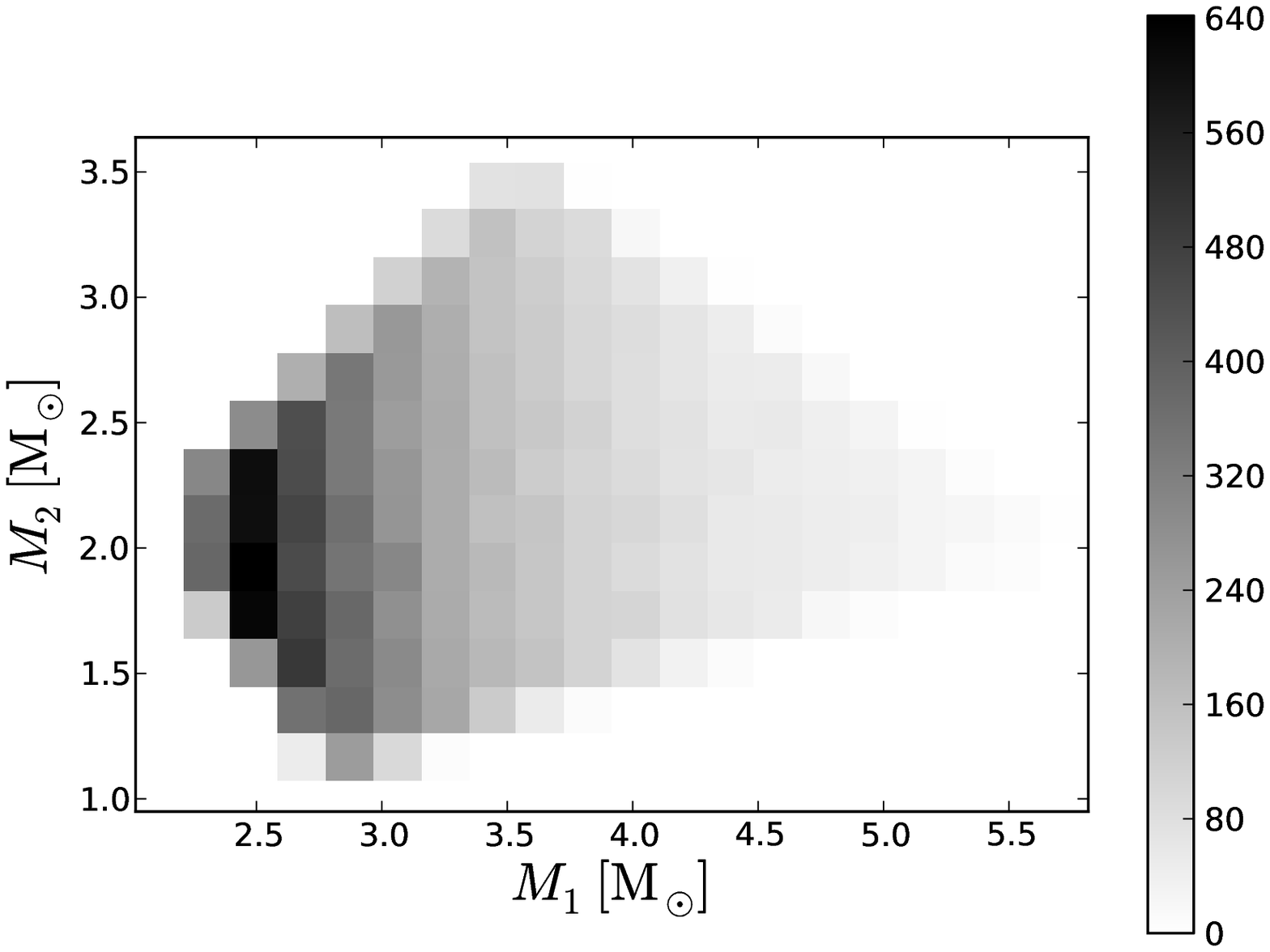}
% \vspace*{-1.0 cm}
        \caption{Density of number of systems in the initial masses plane (the $M_1-M_2$ plane).
        The scale is the number of systems in each square. }
           \label{fig:figdenrat12}
\end{center}
\end{figure}

In Figure \ref{fig:figden12} we show the density of potential SN Ia progenitors in our model
in the $M_{\rm WD1}-M_{\rm core}$ plane. In Figure \ref{fig:figdenrat12} the density in the initial mass of
the binary components plane is given for systems where $M_{\rm WD1} < M_{\rm core}$.
By definition, all systems have $M_1 > M_2$.
However, in our modelling of the CD scenario most, but definitely not all,
potential SN Ia progenitors have $M_{\rm WD1} < M_{\rm core}$.
As expected, the core mass is not much larger than the primary WD mass in the merging systems.
% FFFFFFFFFFFFFFFFFFFFFFFFFFFFFFFFFFFFFFFFFFFFFFFFFFFFFF
% FFFFFFFFFFFFFFFFFFFFFFFFFFFFFFFFFFFFFFFFFFFFFFFFFFFFFF
% FFFFFFFFFFFFFFFFFFFFFFFFFFFFFFFFFFFFFFFFFFFFFFFFFFFFFF
% FFFFFFFFFFFFFFFFFFFFFFFFFFFFFFFFFFFFFFFFFFFFFFFFFFFFFF

\begin{figure}
%  \centering
\hspace*{3.0 cm}
 \includegraphics[scale=0.4]{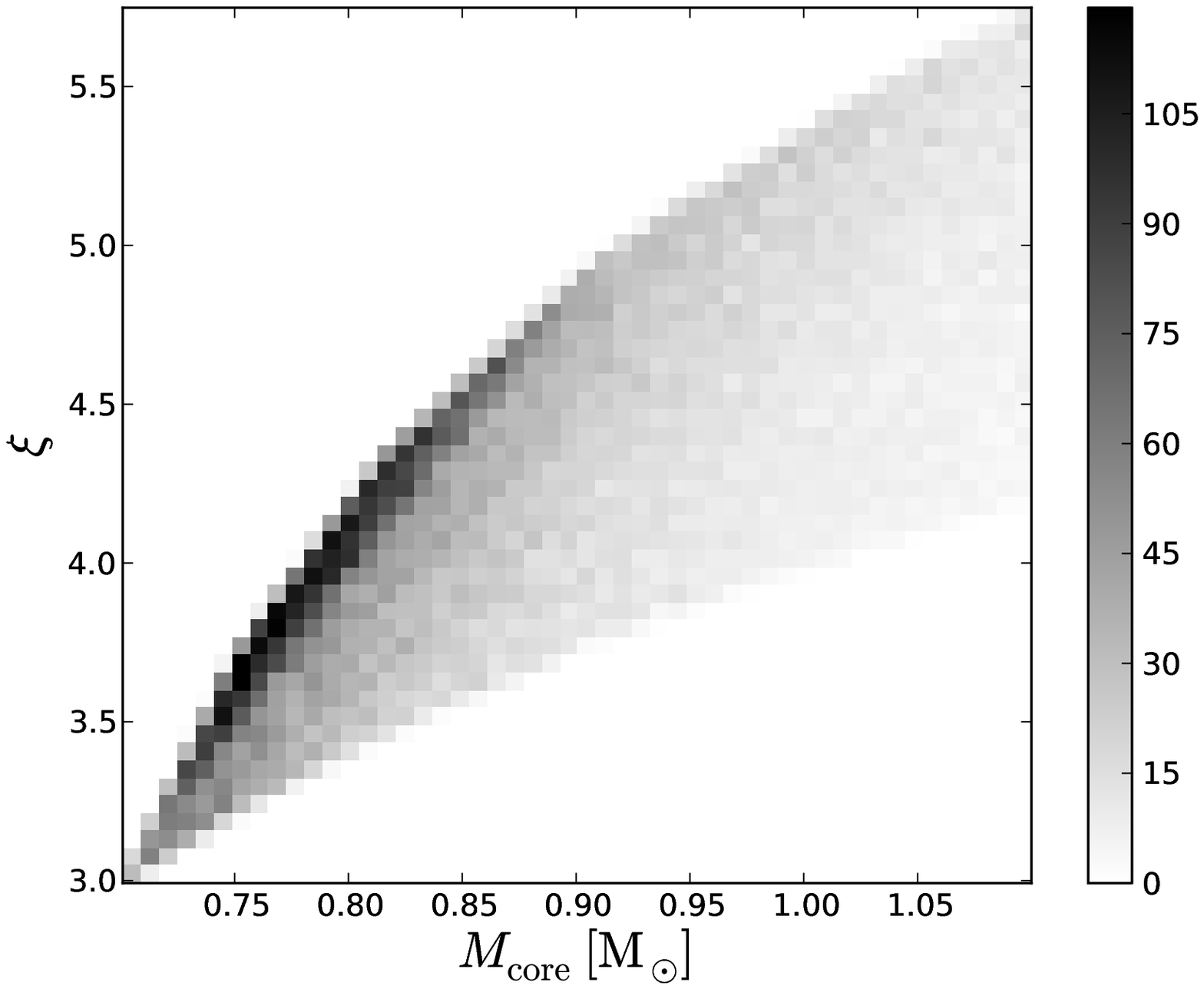}
\vspace*{1.0 cm}
 \\
\hspace*{3.0 cm}
\vspace*{1.0 cm}
%\hspace*{-2.0 cm}
 \includegraphics[scale=0.4]{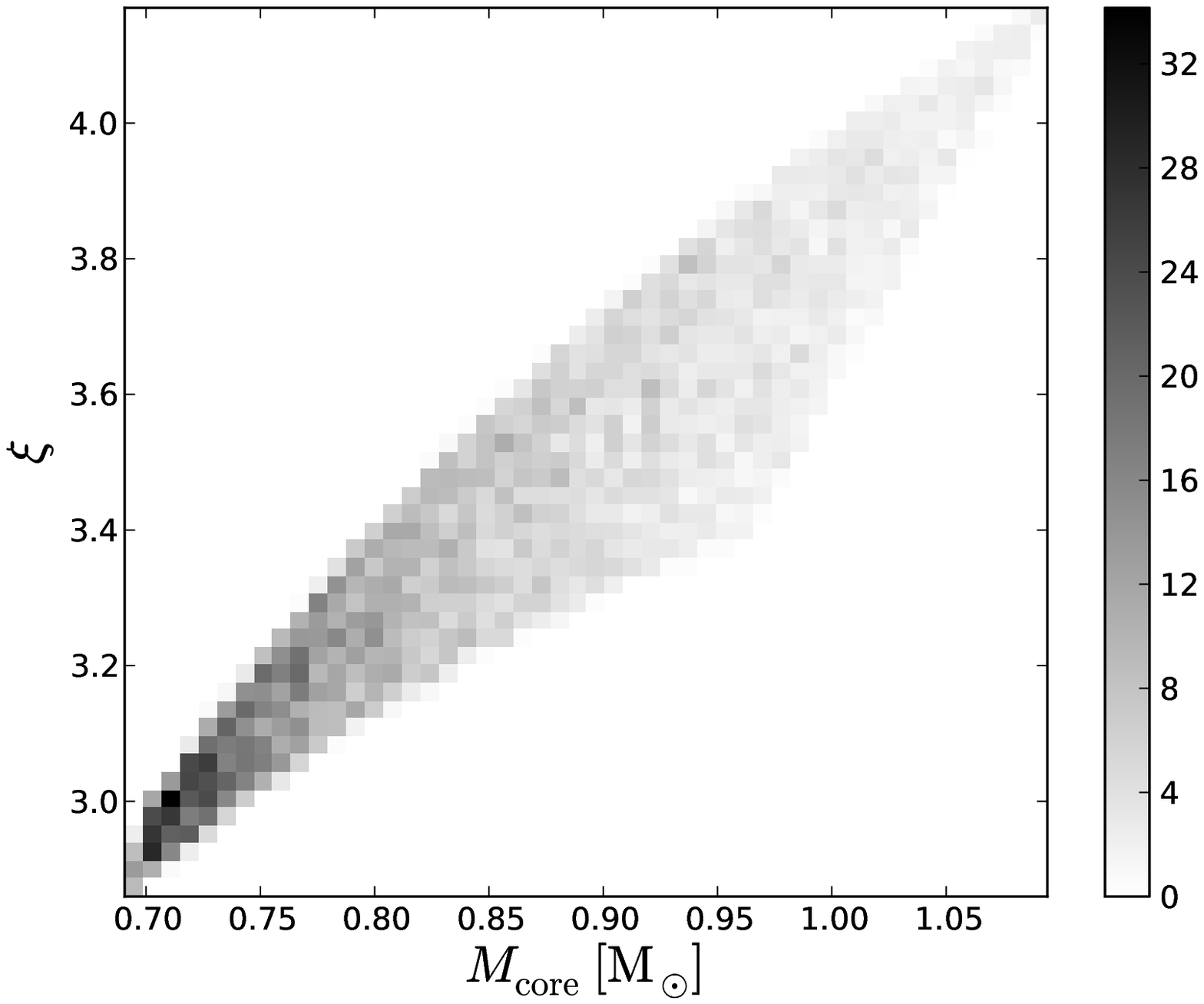}
\caption{The ratio of secondary envelope mass to primary WD mass, $\xi \equiv (M_{\rm 2new} - M_{\rm core})/M_{\rm WD1}$,
as function of $M_{\rm core}$ for $\eta=0.8$.
Upper panel: As with most figures in this subsection, only systems with $M_{\rm WD1}< M_{\rm core}$ are included.
Lower panel: Only systems with $M_{\rm WD1} >  M_{\rm core}$ are included.
High values of $\xi \ga 3-4$ favor merger during the CE phase \citep{Soker2012}.
$M_{2 \rm new}$ is the mass of the secondary after the mass transfer episode from the primary star according to equation (\ref{eq:eta1}).  }
      \label{fig:figxi1}
\end{figure}

In Figure \ref{fig:figxi1} we show the ratio of the envelope mass to the WD mass, $\xi \equiv (M_{\rm 2new} - M_{\rm core})/M_{\rm WD1}$,
as function of core mass, for cases with $M_{\rm WD1}< M_{\rm core}$ (upper panel) and $M_{\rm WD1} >  M_{\rm core}$ (lower panel).
Neglecting mass loss by the secondary star before the onset of the CE phase, $\xi$ is the ratio of secondary envelope mass to
the primary stellar remnant mass.
High values of $\xi \ga 3-4$ favor merger during the CE phase \citep{Soker2012}.
It is evident that systems for which $M_{\rm WD1} < M_{\rm core}$ are more likely to merge.
That the hot core is more massive than the cold WD companion makes
a difference in the  merger process relative to the process of merger of two cold WDs.
% FFFFFFFFFFFFFFFFFFFFFFFFFFFFFFFFFFFFFFFFFFFFFFFF
% FFFFFFFFFFFFFFFFFFFFFFFFFFFFFFFFFFFFFFFFFFFFFFFF

% ==========================================================
\section{DISCUSSION AND SUMMARY}
\label{sec:summary}
% ==========================================================

 In estimating the number of SN Ia progenitors in the core-degenerate (CD) scenario we took a different approach than
 previous population synthesis studies in treating the common envelope (CE) phase.
We assumed that most CE processes where the mass of the white dwarf remnant of the primary (initially more massive) star is less that the
mass of the core,  $M_{\rm WD1} <  M_{\rm core}$, end in mergers.
These systems have a large envelope to WD mass ratio, $\xi \ga 3.5$, as can be seen in the upper panel of Figure \ref{fig:figxi1}.
Such systems have a high probability for WD-core merger \citep{Soker2012}.
This approach avoids using the highly uncertain and controversial $\alpha$-CE parameter in estimating the final orbital separation.

We required that the binary interaction is strong during the late AGB phase of the secondary star, but not during its RGB phase.
We then used the ratio of maximum radii on these two phases, as presented in Figure \ref{fig:fig1}, to find the probability for each
binary system to be in the relevant orbital separation range. The probability is given in equation (\ref{eq:xi1}), where we took the initial orbital separation
of the binary system population to span 4.5 orders of magnitudes.
This simple treatment avoids using the tidal interaction expression that has some uncertainties.
For having a large ratio of AGB maximum radius to RGB maximum radius the initial mass of the star should be $M_{\rm MS} > 2.3 M_\odot$.
Such stars also form WDs of masses that are about half the Chandrasekhar critical mass, such that the merger product of
two such stars reaches the critical mass. This coincidence is discussed in section \ref{sec:coincident}.

These steps, that are described in section \ref{sec:assumptions}, leave us with two parameters. One is the mass transfer parameter from the
primary to the secondary, $\eta$, as given by equation \ref{eq:eta1}, and the other is the binary fraction $f_B$.
The likely values for these are $\eta \simeq 0.9$ and $f_b \simeq 0.65$. We take a conservative approach and use $\eta=0.8$ and $f_b=0.5-0.65$.
The value of $f_B=0.5$ is used by \cite{Nelemansetal2012} in their comparison of different studies.
The criteria for a system to be counted as a SN Ia progenitor are listed in section \ref{subsec:counting}.

The results are given in Table \ref{tab:Table1}. The rows without a superscript $\ast$ include all core-WD systems entering a CE phase, while the rows marked
with a superscript $\ast$ include only systems with $M_{\rm WD1} <  M_{\rm core}$.
The latter cases are more likely to merge and are the systems considered here.
These numbers should be taken very cautiously at this preliminary stage.
As can be seen from Table \ref{tab:Table1}, our very simple approach matches rate of SN Ia deduced from observations.
In section \ref{subsec:number} we also list processes we have neglected, some that will increase and other that will decrease the numbers found here.
Overall, despite the large uncertainties and crude estimate, we find it very satisfactory that the simplest population synthesis we could
think of yields SN Ia average rate in the CD scenario very close to observations.
This match must be reexamined with more sophisticated population synthesis calculations that take mergers into account and where
onset of the CE phase is reconsidered.

\cite{Nelemansetal2012} summarize the results of population synthesis studies of the double-degenerate (DD) scenario from six groups.
The results span the range of $0.2-0.6$ SN Ia per $1000 M_\odot$ for $f_B=0.5$. These numbers are a factor of 2-10 smaller than our results.
We think that two processes combine to give these lower numbers.
First is the CE-$\alpha$ prescription they use.
Some systems in their calculations end at a too large orbital separation to become a SN Ia.
The second one has to do with the treatment of the tidal interaction and mass transfer during the RGB and AGB phases.
The tidal interaction and the formation of a CE phase when the secondary reach the upper AGB might be more significant than what is usually assumed
in population synthesis studies.
This is because of the unstable nature of massive AGB stars, as all secondaries are in our calculations. Such instabilities are likely
to increase the number of WD that enter the CE phase with the secondary.
We call for people who follow the RGB and AGB phases in the simulations to reconsider the tidal interaction when the AGB reaches the upper AGB,
and alow for stronger interaction due to instabilities. Such instabilities might increase the orbital separation from where the WD remnant of the
primary star enters the CE phase.
We intend to study the effects of AGB instabilities in a future paper.

Future studies of the CD scenario will have to explore the exact nature and outcome of the merger process of a hot core with a cooler WD with a similar mass
$0.7 M_{\rm core} \la M_{\rm WD1} \la 1.2 M_{\rm core}$, as this is a major open question in the CD scenario.
Three-dimensional simulations of this process are highly desired before the CD scenario can stand on a more solid ground.

On the observational side, if the CD scenario does indeed account for a substantial fraction of SN Ia, then rapidly rotating super-Chandrasekhar WDs should be found.
Claims for massive rapidly rotating and strongly magnetized WDs have already been made,
e.g. see discussion by \citet{Malheiro2012} and \citet{Wickramasinghe2000}.
Out of the 16 magnetic WDs with well determined mass reported by \citet{Wickramasinghe2000}, two have
masses close to the Chandrasekhar limit, $M>1.3 M_\odot$.
Recent studies also show that magnetic WDs on average are more massive than non-magnetic WDs (e.g., \citealt{Vanlandingham2005}),
in particular for $M_{\rm WD} > 1.3 M_\odot$ \citep{Nalezyty2004}.
WD with a long time delay to explosion should be found in the mass range $\sim 1.4-1.5 M_\odot$
{{{ { (Jorge Rueda, 2012, private communication). } }}}
More massive merger remnants are expected to explode in
star forming regions, and in some cases even have massive hydrogen-rich nebula around them \citep{Sokeretal2012}.
For further discussion on the expected fraction of massive WDs see \cite{IlkovSoker2012}.
There are also theoretical studies of super-Chandrasekhar WDs stabilized by strong magnetic field (e.g., \citealt{Kundu2012}) that should be extended to
systems relevant to the CD scenario.

%%%   ++++++++++++++
%%%   In \cite{Sokeretal2012} a population synthesis of a rare route of the CD scenario where very massive WD merges with a very massive
%%%   ++++++++++++++
%%%   Our Monte Carlo simulations show that the frequency of these systems with the conditions listed above is
%%%   $0.002$ per $1000 M_{\sun}$ stars formed.
%%%   If we relax the assumption on the core+WD mass and the envelope mass to
%%%   $M_{\rm core} + M_{\rm WD} \ga 1.8 M_{\sun}$ (instead of $M_{\rm core} + M_{\rm WD} \ga 2 M_{\sun}$),
%%%   and $M_{\rm env} \ga 0.5M_{\sun}$ (instead of $M_{\rm env} \ga 4M_{\sun}$), respectively,
%%%   we find $0.016$ systems to obey the condition per $1000 M_{\sun}$ stars formed.
%%%   The later envelope mass might be appropriate to SN~2005gj.
%%%   We conclude that the CD scenario can satisfactorily explain all the observed properties of PTF~11kx and, moreover,
%%%   that the number of expected progenitors with massive CSM is consistent with the observations.
%%%   +++++++++++

We thank Xiangcun Meng
{{{ {and Ashley Ruiter} }}} for helpful comments.
This research was supported by the Asher Fund for Space Research at the Technion, and the Israel Science foundation.

% %%%%%%%%%%%%%%%%%%%%%%%%%%%%%%%%%%%%%%%%%%%%%%%%%%%%%%%%%%%%%%%%%%%%%%%%%%%%%%%%%%%%%
% %%%%%%%%%%%%References
% %%%%%%%%%%%%%%%%%%%%%%%%%%%%%%%%%%%%%%%%%%%%%%%%%%%%%%%%%%%%%%%%%%%%%%%%%%%%%%%%%%%%%

\label{lastpage}

\end{document}